\documentclass[12pt]{article}

\usepackage{amsfonts,amssymb,amsmath}
\usepackage{tikz}
\usepackage{amsthm}
\usepackage[font={small}]{caption}
\usepackage{subcaption}

\newtheorem{prop}{Proposition}

\usetikzlibrary{positioning,calc}
\usetikzlibrary{decorations.markings}

\newcommand{\ra}{{\rightarrow}}

\newcommand{\ZZ}{{\mathbb Z}}

\newcommand{\CC}{{\mathbb C}}

\newcommand{\Vect}{{\rm Vect}}

\newcommand{\Aut}{{\rm Aut}}

\newcommand{\Tr}{{\rm Tr}}

\newcommand{\cA}{{\mathcal A}}

\newcommand{\z}{{\vec z}}

\newcommand{\C}{{\mathcal C}}

\newcommand{\cT}{{\mathcal T}}

\title{Equivariant Topological Quantum Field Theory and Symmetry Protected Topological Phases}

\author{Anton Kapustin\footnote{On leave of absence from California Institute of Technology} \\ \it Simons Center for Geometry and Physics, Stony Brook, NY \\ Alex Turzillo \\ \it California Institute of Technology, Pasadena, CA}

\begin{document}

\maketitle

\abstract{Short-range entangled topological phases of matter are closely connected to Topological Quantum Field Theory. We use this connection to classify Symmetry Protected Topological Phases in low dimensions, including the case when the symmetry involves time-reversal. To accomplish this, we generalize Turaev's description of equivariant TQFT to the unoriented case. We show that invertible unoriented equivariant TQFTs in one or less spatial dimensions are classified by twisted group cohomology, in agreement with the group cohomology proposal of Chen, Gu, Liu and Wen. We also show that invertible oriented equivariant TQFTs in spatial dimension two or less are classified by ordinary group cohomology. }

\newpage

\section{Introduction and overview}

Recently the problem of classifying gapped phases of matter whose ground state is short-range entangled (SRE phases) received a lot of attention.\footnote{There are at least two  slightly different definitions of short-range entanglement, see \cite{ChenGuWen} and \cite{Kitaevtalk}. The difference between them is discussed in section 2.4.} SRE phases can be divided into two broad classes, bosonic and fermionic, depending on whether the fundamental degrees of freedom are bosons or fermions. The bosonic SRE phases are in many ways simpler, and there has been a substantial progress in their classification. In particular, it has been proposed in \cite{groupcohomology} that $D$-dimensional bosonic SRE phases with a finite internal symmetry $G$ are classified by the abelian group $H^{D+1}(BG,U(1))$. Here $BG$ is the classifying space of $G$, and $D$ is the dimension of space (thus the dimension of space-time is $D+1$).  Later it was noticed that some SRE phases in spatial dimension $3$ are not captured  by the group cohomology classification \cite{beyondGC}, and it was proposed by one of the authors that the group cohomology classification can be improved by replacing ordinary cohomology of $BG$ with a particular generalized cohomology theory (the stable cobordism) \cite{cobordism}. Other generalized cohomology theories have also been proposed as candidates for the classification scheme \cite{Freed, Kitaevtalk}, and it appears that the answer might depend on the detailed assumptions about the properties of SRE phases . But for $D\leq 2$ all classification schemes agree, and in fact in the Hamiltonian approach one can use the matrix product representation of SRE states to prove that $D=1$ bosonic SRE phases are classified by $H^2(BG,U(1))$ \cite{ChenGuWen2}. The $D=1$ fermionic  SRE phases have also been classified \cite{KitaevFidkowski}. The $D=0$ case is even simpler. 

One promising avenue for extending these results to higher dimensions is via equivariant Topological Quantum Field Theory. It is a very attractive conjecture that a large class of gapped phases is described at large scales by a TQFT.\footnote{There are some exceptions to this rule, due to the existence of phases with a non-vanishing thermal Hall conductivity. These exceptions only occur when $D=2\, {\rm mod}\, 4$, because only in these dimensions there exist gravitational Chern-Simons terms.} Both gapped phases and TQFTs can be tensored, and this operation makes both sets into commutative semigroups (sets with an associative and commutative binary operation). Both semigroups have a neutral element ${\bf 1}$ corresponding to the trivial gapped phase or a TQFT. An element $\Phi$ of a semigroup is said to be invertible if there exists an element $\bar \Phi$ such that $\Phi\circ \bar\Phi=\bar\Phi\circ\Phi={\bf 1}$. Thus it makes sense to talk about invertible gapped phases and  invertible TQFTs. According to one of the definitions of SRE phases \cite{Kitaevtalk}, an invertible gapped phase is the same as an SRE phase.  Thus, if one believes into the correspondence between  gapped phases and TQFTs, the classification of SRE phases is reduced to the classification of invertible TQFTs. More generally, SRE phases with a symmetry $G$ correspond to invertible $G$-equivariant TQFTs. 

While classifying TQFTs in $D>1$ is unrealistic, classifying invertible ones is much simpler. In fact, using the known algebraic description of equivariant TQFTs in $D=0,1$ and $2$, it is easy to check that in these dimensions invertible $G$-equivariant TQFTs are classified by $H^{D+1}(BG,U(1))$, provided the group $G$ does not act on space-time (see Section 2). But if some elements of $G$ involve time-reversal, the problem is more complicated. From the TQFT viewpoint, time-reversal symmetry means that the theory can be defined on unorientable space-times. The difficulty is that an algebraic description of unoriented equivariant TQFTs is not known even in low dimensions. The main goal of this paper is to provide such an algebraic description in $D=0$ and $D=1$ and to show that invertible equivariant TQFTs are classified by twisted group cohomology $H^{D+1}(BG,U(1)_\rho)$, where $\rho: G\ra\ZZ_2$ is a homomorphism which tells us which elements of $G$ are time-reversing and which are not. This agrees with the proposal of \cite{groupcohomology}. It is likely that this method can be extended to $D=2$. In higher dimensions an algebraic description of general TQFTs is prohibitively complicated, and this approach to classifying SRE phases becomes impractical. Note that equivariant TQFTs which are not necessarily invertible are interesting in their own right, as they describe Symmetry Enhanced Topological (SET) phases. 

In Section 2 we deal with  the case of a finite symmetry $G$ which acts trivially on space-time. We recall algebraic descriptions of  oriented equivariant TQFTs in $D\leq 2$ and show that invertible equivariant TQFTs are classified by elements of $H^{D+1}(BG,U(1))$. All of this is either trivial ($D=0$) or well-known to experts ($D=1$ and $D=2$). 

In Section 3 we consider unoriented equivariant TQFT in $D=0$ and the corresponding SRE phases with time-reversing symmetries.

In Section 4 we formulate axioms of unoriented equivariant TQFT in $D=1$ by extending Turaev's axioms in the oriented case \cite{Turaev}.  We show how these axioms lead to a generalization of Turaev's $G$-crossed algebra, which we call $\rho$-twisted $G$-crossed algebra. We prove that every $\rho$-twisted $G$-crossed algebra gives rise to an unoriented equivariant TQFT. Finally we show that invertible TQFTs in $D=1$ give rise to $\rho$-twisted 2-cocycles on $BG$, and that conversely to every element of $H^2(BG,U(1)_\rho)$  one can associate a $\rho$-twisted $G$-crossed algebra which is unique up to isotopy.

It would be interesting to give an algebraic description of $D=2$ unoriented equivariant TQFTs and show that in the invertible case they are classified by $H^3(BG,U(1)_\rho)$. The first step is to categorify our algebraic description of $D=1$ unoriented equivariant TQFT by replacing vector spaces with categories, linear maps with functors, and equalities with isomorphisms. The nontrivial part is to find a complete set of coherence conditions between isomorphisms analogous to the pentagon and hexagon conditions in the oriented case which ensure consistency under gluing. 

A.K. would like to thank V. Ostrik for helpful discussions. 
The work of A.K. was supported by the Simons Foundation. The work of  A. T.  was supported in part by the DOE grant  DE-FG02-92ER40701.

\section{Oriented equivariant TQFT and SRE phases with an internal symmetry}

\subsection{$D=0$}

A  $D=0$ TQFT is ordinary quantum mechanics with zero Hamiltonian and is completely determined by its space of states (a finite-dimensional complex vector space $V$). Equivariant TQFT is merely a vector space $V$ with an action of $G$. Since $G$ is finite, this representation is unitarizable (unitary for a suitable choice of inner product on $V$). The trivial equivariant TQFT corresponds to $V=\CC$ with a trivial action of $G$. Equivariant TQFTs which are invertible with respect to the tensor product are one-dimensional representations of $G$, i.e., elements of $H^1(BG,\CC^*)\simeq H^1(BG,U(1))$. 

\subsection{$D=1$} 

 $D=1$ TQFTs  are in one-one correspondence with commutative Frobenius algebras \cite{Atiyah} (see \cite{MooreSegal} for a nice exposition, including various generalizations). The vector space $\cA$ underlying the algebra  is the space of states of the TQFT on a circle. The state-operator correspondence identifies $\cA$ with the space of local operators, which is clearly a commutative algebra. The Frobenius structure is a non-degenerate bilinear inner product
 $$
 \eta (a,b)\in\CC,\quad a,b\in \cA,
 $$ 
 satisfying $\eta(ab,c)=\eta(a,bc)$. It is a combination of the usual sesquilinear Hilbert space inner product and the anti-linear CPT transformation:
 $$
 \eta(a,b)= ({\rm CPT} a, b).
 $$

A trivial $D=1$ TQFT has $\cA\simeq \CC$ and $\eta(1,1)=1$. An invertible TQFT has $\cA\simeq \CC$, and thus is completely determined by $\eta(1,1)\in\CC^*=\CC\backslash \{0\}$. But if we are interested only in classifying TQFTs up to isotopy (i.e. up to continuous deformations), then all these TQFTs can be identified (since $\pi_0(\CC^*)$ is trivial). If we identify  invertible TQFTs and SRE phases, this means that in the absence of symmetry there are no nontrivial $D=1$ SRE phases.

To incorporate a symmetry $G$, we need to consider $G$-equivariant $D=1$ TQFTs. $G$-equivariance means that we can couple the theory to an arbitrary $G$-bundle. The precise definition of equivariant TQFT will be recalled in section 3. For now, we only need the algebraic description of such TQFTs due to Turaev \cite{Turaev}. He defines a $G$-crossed algebra as a finite-dimensional Frobenius algebra $(\cA=\oplus_{g\in G}\cA_g, \eta)$ together with a homomorphism $\alpha: G\ra \Aut\,\cA$ such that 
\begin{gather}
\cA_g\cdot \cA_h\subset \cA_{gh}\ {\rm and}\ 1\in \cA_1. \label{o1}\\
\eta(\cA_g, \cA_h)=0\ {\rm if}\ gh\neq 1. \label{o2}\\
\alpha_h(\cA_g)\subset \cA_{h g h^{-1}}. \label{o3}\\
\alpha\ {\rm preserves}\ \eta \ {\rm and}\ \alpha_h\vert_{\cA_h}={\rm id}. \label{o4}\\
\forall \psi_g\in\cA_g, \psi_h\in\cA_h\ {\rm we\ have}\ \psi_g\cdot\psi_h=\alpha_g(\psi_h)\cdot\psi_g. \label{o5}\\
\begin{split} \forall g\in G\ {\rm  let}\ \xi_i^g\ {\rm and}\ \xi^i_g\ & {\rm  be\ dual\ bases\ in}\ \cA_g\ {\rm  and}\ \cA_{g^{-1}}.\ {\rm Then} \\
\sum_i \alpha_h(\xi_i^g) & \xi^i_g =\sum_j \xi_j^h \alpha_g(\xi^j_h),\ \forall g,h\in G. \label{o6} \end{split}
\end{gather}

Let us make a few remarks about this definition. $\cA_g$ is the $g$-twisted sector of the space of states on a circle. $\alpha_h$ describes the action of $G$ on the space of states. If $G$ is abelian, it acts on each twisted sector separately, but in general it mixes different twisted sectors. The penultimate axiom shows that $\cA$ is not commutative, but is twisted-commutative.
The last axiom arises from considering a punctured torus with twists by $g$ and $h$ along the two generators of its fundamental group and computing the corresponding state in two different ways. This axiom, together with the Frobenius condition $\eta(a,bc)=\eta(ab,c)$, implies
$$
\dim \cA_g=\Tr\ \alpha_g\vert_{\cA_1}.
$$
Both sides of this equality compute the partition function of a torus twisted by $g$ along one direction and by $1$ along the other direction. On the left-hand side, the direction twisted by $g$ is regarded as space and the direction twisted by $1$ is regarded as time. On the right-hand side, it is the other way around.
Since the right-hand side is a character of a finite group $G$, we get an inequality $0<\dim\cA_g\leq \dim \cA_1$. That is, twisting by $g$ cannot increase the number of states.

In particular, let us consider an invertible $G$-equivariant TQFT. Then $\cA_1\simeq \CC$, and therefore $\cA_g\simeq \CC$ for all $g\in G$. If we choose a basis vector $\ell_h$ in each $\cA_h$, we see that the algebra structure is given by a collection of complex numbers $\theta(g,h)$ such that
$$
\ell_g\cdot \ell_h=b(g,h)\cdot \ell_{gh}.
$$
Twisted commutativity of $\cA$ implies that $b(g,h)$ is nonzero for all $g,h$ and fixes $\alpha_h$ in terms of $b$. Associativity of multiplication implies that $b$ is a 2-cocycle, and changing a basis in $\cA_g$ changes it by a coboundary. The rest of the axioms are easily checked. With $b$ fixed, the only freedom left is the choice of the inner product $\eta$; all such choices lead to isotopic TQFTs, which means that isotopy classes of invertible oriented equivariant $D=1$ TQFTs are classified by $[b]\in H^2(BG,\CC^*)\simeq H^2(BG,U(1))$. This result has been proved in \cite{Turaev}.


\subsection{$D=2$}

When studying oriented $D=2$ TQFTs one usually assumes that the space of local operators (i.e. the vector space attached to $S^2$) is one-dimensional, and thus the algebra of local operators is isomorphic to $\CC$. If one is interested only in unitarizable TQFTs, one does not loose much by focusing on this special case. Indeed, it is easy to show that if the TQFT is unitarizable (i.e. the bilinear inner product arises from a Hermitian inner product and an ant-linear CPT symmetry), then the algebra of local operators is semi-simple. It is also commutative, and therefore isomorphic to a sum of several copies of $\CC$.  The generators of this algebra label different superselection sectors, and one might as well focus on a single sector where all but one generator act trivially. The argument applies equally well for all $D>0$, but in $D=1$ it is traditional to allow the algebra of local operators to be non-semi-simple, in view of string theory applications which require one to consider non-unitary TQFTs. 

We are mostly interested in unitarizable TQFTs, and therefore in this section we assume that the space of local operators is $\CC$. Such oriented $D=2$ TQFTs are described by   modular tensor categories with vanishing central charge $c\in\ZZ/8$ \cite{MooreSeiberg,BK}. (If the central charge is nonzero, one gets a framed $D=2$ TQFT). The data of a modular tensor category attach a vector space to every closed oriented 2-manifold, and a map of vector spaces to every oriented bordism between such 2-manifolds. Similarly, oriented equivariant $D=2$ TQFT is described by a $G$-modular category \cite{Turaev2,TuraevVir}. Its definition is a categorification of the notion of $G$-crossed algebra. In particular, for every $g\in G$ one has a category $\C_g$, and a bi-functor $\C_g\times \C_h\ra \C_{gh}$ satisfying the associativity constraint. The data of a $G$-modular category attach a vector space to every closed oriented 2-manifold with a $G$-bundle and a trivialization at a base point, and a map of vector spaces for every oriented $G$-bordism between such 2-manifolds (i.e. to every oriented 3-manifold with a $G$-bundle which ``interpolates'' between the two oriented 2-manifolds with $G$-bundles). Objects of the category $\C_g$ represent quasi-particles in the $g$-twisted sector. 

An invertible oriented equivariant $D=2$ TQFT is described by a $G$-modular  category with $\C_1\simeq \Vect$, where $\Vect$ is the category of finite-dimensional vector spaces. This condition ensures that for the trivial $G$-bundle the vector space attached to any oriented  2-manifold is one-dimensional. If the TQFT describes a gapped phase, this means that the space of ground states is non-degenerate for any topology. This is a hallmark of an SRE phase. 

From $\C_1\simeq \Vect$ one can deduce that $\C_g\simeq \Vect$ for all $g\in G$. Indeed, by the definition of a $G$-modular category \cite{TuraevVir}, $\C_g$ is nonempty for all $g\in G$. Then Prop. 4.58 in \cite{Drinfeldetal} implies that $\C_g\simeq \Vect$. As a consequence, the vector space attached to any 2-manifold with any $G$-bundle is one-dimensional. That is, there is no ground-state degeneracy even after twisting by an arbitrary $G$-bundle. 

Finally, Prop. 4.61 in \cite{Drinfeldetal} tells us that in the invertible case $\C$ is entirely determined by an element of $H^3(BG,\CC^*)\simeq H^3(BG,U(1))$. 
This agrees with the group cohomology proposal which says that $D=2$ bosonic SRE phases with symmetry $G$ are classified by elements of $H^3(BG,U(1))$. 

\subsection{On the definition of SRE phases with symmetry $G$}

Kitaev \cite{Kitaevtalk} proposed to define SRE phases as invertible gapped phases. That is, a gapped phase $\Phi$ is an SRE phase if there exists another gapped phase $\bar \Phi$ such that $\Phi\otimes\bar \Phi$ can be deformed to the trivial gapped phase without closing the gap. This definition ensures that SRE phases form an abelian group. But if an SRE phase $\Phi$ has a symmetry $G$, a problem arises: it is not clear from this definition whether $\bar \Phi$ can be chosen symmetric,  and whether  the deformation of $\Phi\otimes\bar\Phi$  to the trivial phase can be chosen so that it does not break the symmetry. In particular, suppose we define a Symmetry Protected Topological (SPT) phase as an SRE phase with symmetry $G$ which is trivial if we ignore symmetry, but cannot be deformed to the trivial gapped phase if $G$ is required to be preserved \cite{ChenGuWen}. Then it is not clear whether SPT phases with a fixed symmetry $G$ form an abelian group. In the TQFT world, an analogous question can be formulated as follows. Consider a forgetful map $\Psi$ from the set of $G$-equivariant TQFTs to the set of arbitrary TQFTs. Let $\cT$ be a $G$-equivariant TQFT such that $\Psi(\cT)$ is invertible. Is it true that $\cT$ is invertible? The discussion in this section implies that this is true for $D\leq 2$ and if $G$ does not involve time-reversal.

\section{Unoriented equivariant $D=0$ TQFT }

In the $D=0$ case, the homomorphism $\rho:G\ra \ZZ_2$ tells us whether a particular element $g$ reverses the direction of time.  
Our goal is to show that invertible unoriented equivariant TQFTs in $D=0$ spatial dimensions are classified by the twisted cohomology group $H^1(BG,U(1)_\rho)$. Recall that a $\rho$-twisted 1-cochain on $BG$ is the same as a function $\phi:G\ra U(1)$ satisfying
$$
\phi(gh)=\phi(g)\phi(h)^{\rho(g)}.
$$
Here and below we identify $\ZZ_2$ with $\{1,-1\}$, and thus $\rho(g)=-1$ if $g$ is time-reversing and $\rho(g)=1$ otherwise.
Two twisted cochains $\phi(g)$ and $\psi(g)$ are regarded as equivalent (i.e. cohomologous) if there exists $\mu\in U(1)$ such that for all $g\in G$ we have 
$$
\psi(g)=\mu^{\rho(g)-1}\phi(g)= \left\{ \begin{array}{lr} \phi(g), & \rho(g)=1,\\ \mu^{-2}\phi(g), & \rho(g)=-1.\end{array}\right .
$$

A $D=0$ TQFT associates a complex vector space $V$ to a point. To each $g\in G$ it associates an operator 
$$
\Lambda(g): V\ra V.
$$
where $\Lambda(g)$ is linear if $\rho(g)=1$ and anti-linear if $\rho(g)=-1$. After choosing a basis in $V$, we can attach to every $\Lambda(g)$ a complex non-degenerate matrix $M(g)$, by letting
$$
\Lambda(g)=\left\{ \begin{array}{lr} M(g), & \rho(g)=1,\\ M(g) K, & \rho(g)=-1.\end{array}\right .
$$
Here $K: V\ra V$ is an operator which complex-conjugates the coordinates of a vector in the chosen basis. The matrices $M(g)$ do not form a complex representations of $G$, rather \cite{Weyl}:
$$
M(g_1g_2)=\left\{\begin{array}{lr} M(g_1) M(g_2), & \rho(g_1)=1, \\ M(g_1) M(g_2)^*, & \rho(g_1)=-1.\end{array}\right .
$$
In the invertible case $V\simeq \CC$ the matrices $M(g)$ become elements of $\CC^*$, and the above equation becomes precisely the twisted cocycle condition for the $\CC^*$-valued 1-cochain $M(g)$, where $\ZZ_2$ acts on $\CC^*$ by complex conjugation. 

We should also investigate the effect of a change of basis in $V$. In the invertible case, if we replace the basis element $\ell\in V$ by $\lambda^{-1} \ell$, $\lambda\in\CC^*$, the function $M(g)$ transforms as follows:
$$
M(g)\mapsto \left\{ \begin{array}{lr} M(g), & \rho(g)=0,\\ \lambda^{-1}\lambda^* M(g),&  \rho(g)=1.\end{array} \right .
$$
This is precisely the shift of the twisted 1-cocycle $M(g)$ by a twisted coboundary. Thus equivalence classes of invertible unoriented equivariant $D=0$ TQFTs are classified by elements of $H^1(BG,\CC^*_\rho)\simeq H^1(BG,U(1)_\rho)$.

\section{Unoriented equivariant $D=1$ TQFT}

\subsection{Definition of unoriented equivariant TQFT}

For $D>0$ we can avoid anti-linear operators by interpreting the orientation-reversing symmetry as a parity symmetry ($P$ or $CP$). Since $CPT$ is a symmetry of any local unitary QFT, we do not loose generality by doing this. Thus $\rho(g)=-1$ if $g$ reverses spatial orientation, $\rho(g)=1$ otherwise.

At first we will try to be as general as possible and do not fix the spatial dimension $D$. We consider a finite group $G$ together with a homomorphism $\rho:G\ra\ZZ_2$. The kernel of $\rho$ will be denoted $G_0$. For any manifold $X$ we will denote by $o(X)$ its orientation bundle. Any TQFT is defined as a functor from a geometric source category with a symmetric monoidal structure to the category of finite-dimensional vector spaces $\Vect$ (or more generally, to a symmetric monoidal category). 

In the case of equivariant TQFT based on the pair $(G,\rho)$ he source category $\C$ is defined as follows. An object of $\C$ is a closed $D$-manifold $M$, a base point for every connected component of $M$, a $G$-bundle $E$ over $M$, a trivialization of $G$ at every base point, and a trivialization of $o(M)\otimes \rho(E)$ everywhere on $M$. The last datum expresses the fact that $\rho(E)$ is isomorphic to the orientation bundle of $X$. A morphism of $\C$ is an isomorphism class of a $D+1$-dimensional bordism $N$ equipped with a $G$-bundle $E$ and a trivialization of $o(N)\otimes \rho(E)$, with every connected component of the boundary given a base point and a trivialization of $E$ there. Two such bundles are said to be isomorphic if they are related by a bundle map that is an homeomorphism of the total space, covers a homeomorphism of the base space, and preserves the trivialization and boundary data. There is also a decomposition of the boundary into two disjoint parts, corresponding to the source and target of the morphism. Composition of morphisms is obvious. The symmetric monoidal structure arises from the operation of disjoint union. 


Let us now specialize to the case $D=1$. In this case the definition can be simplified, because all 1d manifolds are orientable. Since we are given trivializations of $E$ at all base points, as well a trivialization of $o(M)\otimes \rho(E)$, we also have a trivialization of $o(M)$ at all base points. But since $M$ is orientable, this means that we are given a trivialization of $o(M)$ everywhere, i.e. an orientation. Then $\rho(E)$ is also trivialized everywhere, and the $G$-bundle reduces to a $G_0$-bundle. Thus the objects for $\C$ are exactly the same as in the oriented equivariant TQFT with symmetry group $G_0$. Morphisms are different however, for example because unorientable bordisms are now allowed. Moreover, even when bordisms are orientable, they are not given an orientation.  More precisely, if the boundary of a bordism is connected, there is a base point with an orientation on it, and one can use this to extend orientation to the whole $N$. But if more than one base point is present, there is no guarantee that orientations so obtained agree between each other. This will be discussed in more detail below.

\subsection{Algebraic description for $D=1$}

From the above definition we extract  the following algebraic data. First of all, let $M=S^1$. As remarked above, $S^1$ is actually oriented, and the structure group $G$ is reduced to $G_0$. Thus unoriented equivariant TQFT assigns a vector space $\cA_g$ to every $g\in G_0$. 

Now consider a cylinder regarded as a bordism from $S^1$ to $S^1$. It has two marked points on the boundaries which we call $p_-$ and $p_+$ (source and target). A $G$-bundle over a cylinder trivialized over $p_-$ is determined by the holonomy around the source $S^1$ and thus is labeled by an element $g\in G$. We are also given a trivialization at $p_+$, and the holonomy along a path from $p_-$ to $p_+$ gives a well-defined element  $h\in G$. 
We know that $g\in G_0$, but $h$ can be an arbitrary element of $G$. If $\rho(h)=1$, the two trivializations of $\rho(E)$ obtained from the trivializations of $E$ at $p_-$ and $p_+$ agree. Then, since $o(N)\otimes\rho(E)$ is trivialized everywhere, the orientations at $p_-$ and $p_+$ also agree, and the source and target circles have the same orientation. Thus the source is labeled by $g$, and the target by $hgh^{-1}$, and the cylinder is assigned a map $\alpha_h: \cA_g\ra \cA_{h g h^{-1}}$. Similarly, if $\rho(h)=-1$, the two orientations disagree, and the target is labeled by $h g^{-1} h^{-1}$, while the source is still labeled by $g$.
Such a cylinder is assigned a map $\alpha_h: \cA_g\ra \cA_{h g^{-1} h^{-1}}$. We can summarize both cases by saying that $\alpha_h$ maps $\cA_g$ to $\cA_{h g^{\rho(g)} h^{-1}}.$ Since gluing two cylinders labeled by $(g,h)$ and $(hg^{\rho(g)} h^{-1} ,h')$ using the trivial identification of target and source circles gives a cylinder labeled by $(g,h'h)$, we must have $\alpha_{h'}\circ\alpha_h=\alpha_{h'h}$. In particular, each $\alpha_h$ is invertible.

\begin{figure}[ht]
\centering
\begin{subfigure}{\textwidth}
\centering
\begin{tikzpicture}[cross/.style={path picture={ 
  \draw[black]
(path picture bounding box.south east) -- (path picture bounding box.north west) (path picture bounding box.south west) -- (path picture bounding box.north east);
}}]

\node[draw,circle,cross,label=left:$g$] at (2,2) {};

\draw (1.2,2) arc (180:270:0.8);
\draw[decoration={markings, mark=at position 1 with {\arrow[scale=2]{>}}}, postaction={decorate}] (2,2) arc (0:180:0.4);
\draw[decoration={markings, mark=at position 0.6 with {\arrow[scale=2]{<}}}, postaction={decorate}] (2,2) -- (2,1.2);
\draw [fill] (2,1.2) circle [radius=0.07];

\draw[dotted, decoration={markings, mark=at position 0 with {\arrow[scale=2]{>}}}, postaction={decorate}] (2,2) circle (0.8);
\draw[decoration={markings, mark=at position 0 with {\arrow[scale=2]{>}}}, postaction={decorate}] (2,2) circle (2);
\draw[decoration={markings, mark=at position 0.4 with {\arrow[scale=2]{>}}}, postaction={decorate}] (2,1.2) -- (2,0) node[midway,right]{$h$};

\node[draw,circle,cross] at (7,2) {};

\draw plot [smooth,tension=1] coordinates{(6.2,2) (6.3,1.1) (6.6,0.4) (6.9,0.1) (7,0)};
\draw[decoration={markings, mark=at position 1 with {\arrow[scale=2]{>}}}, postaction={decorate}] (7,2) arc (0:180:0.4);

\draw[decoration={markings, mark=at position 0 with {\arrow[scale=2]{>}}}, postaction={decorate}] (7,2) circle (2) node [label={[label distance=-2.9cm]0:$\cong$}]{};
\draw[decoration={markings, mark=at position 0.4 with {\arrow[scale=2]{<}}}, postaction={decorate}] (7,2) -- (7,0) node[midway,right]{$hgh^{-1}$};

\draw [fill] (2,0) circle [radius=0.07];
\draw [fill] (7,0) circle [radius=0.07];

\end{tikzpicture}
\caption{Axiom \eqref{eqax4} for $h\in G_0$.} \label{fig1a}
\end{subfigure}
\quad
\begin{subfigure}{\textwidth}
\centering
\begin{tikzpicture}[cross/.style={path picture={ 
  \draw[black]
(path picture bounding box.south east) -- (path picture bounding box.north west) (path picture bounding box.south west) -- (path picture bounding box.north east);
}}]

\node[draw,circle,cross,label=left:$g$] at (2,2) {};

\draw (1.2,2) arc (180:270:0.8);
\draw[decoration={markings, mark=at position 1 with {\arrow[scale=2]{>}}}, postaction={decorate}] (2,2) arc (0:180:0.4);
\draw[decoration={markings, mark=at position 0.6 with {\arrow[scale=2]{<}}}, postaction={decorate}] (2,2) -- (2,1.2);
\draw [fill] (2,1.2) circle [radius=0.07];

\draw[dotted, decoration={markings, mark=at position 0 with {\arrow[scale=2]{>}}}, postaction={decorate}] (2,2) circle (0.8);
\draw[decoration={markings, mark=at position 0 with {\arrow[scale=2]{<}}}, postaction={decorate}] (2,2) circle (2);
\draw[decoration={markings, mark=at position 0.4 with {\arrow[scale=2]{>}}}, postaction={decorate}] (2,1.2) -- (2,0) node[midway,right]{$h$};

\node[draw,circle,cross] at (7,2) {};

\draw plot [smooth,tension=1] coordinates{(6.2,2) (6.3,1.1) (6.6,0.4) (6.9,0.1) (7,0)};
\draw[decoration={markings, mark=at position 1 with {\arrow[scale=2]{>}}}, postaction={decorate}] (7,2) arc (0:180:0.4);

\draw[decoration={markings, mark=at position 0 with {\arrow[scale=2]{>}}}, postaction={decorate}] (7,2) circle (2) node [label={[label distance=-2.9cm]0:$\cong$}]{};
\draw[decoration={markings, mark=at position 0.4 with {\arrow[scale=2]{<}}}, postaction={decorate}] (7,2) -- (7,0) node[midway,right]{$hg^{-1}h^{-1}$};

\draw [fill] (2,0) circle [radius=0.07];
\draw [fill] (7,0) circle [radius=0.07];

\end{tikzpicture}
\caption{Axiom \eqref{eqax4} for $h\notin G_0$.} \label{fig1b}
\end{subfigure}
\end{figure}

In general, we note that if $N$ is an orientable bordism, and the paths between base points on different boundary components all lie in $G_0$, the morphism becomes a morphism in the oriented equivariant theory with symmetry group $G_0$. Thus we get all the same algebraic data as in the oriented $G_0$-equivariant theory.  That is, a $G_0$-crossed algebra 
$$\cA=\oplus_{g\in G_0} \cA_g,\quad \eta: \cA\otimes\cA\ra \CC,\quad \alpha: G_0\ra {\rm Aut}\, \cA,$$
satisfying \eqref{o1}-\eqref{o6}. In particular, for $h\in G_0$ the map $\alpha_h$ is an automorphism of $\cA$. On the other hand, for $h\notin G_0$ the map $\alpha_h$ is an anti-automorphism:
\begin{equation}\label{antiaut}\alpha_h(ab)=\alpha_h(b)\alpha_h(a),\quad \forall h\notin G_0, \forall a,b\in \cA.\end{equation}
To see this, we compare the two pants diagram with cylinders attached either to the torso or to the pant legs and note that for $h\notin G_0$ they are related by a reflection rather than the identity homeomorphism.

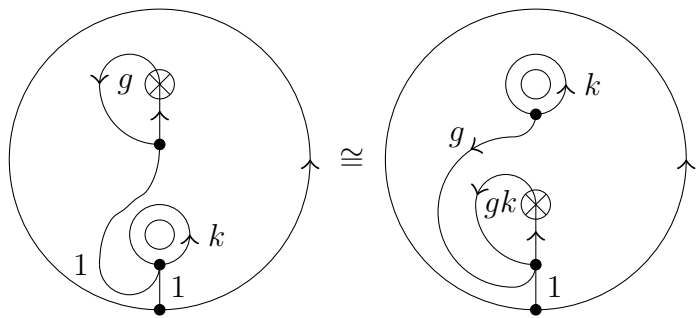
\begin{figure}[ht]
\centering
\begin{tikzpicture}[cross/.style={path picture={ 
  \draw[black]
(path picture bounding box.south east) -- (path picture bounding box.north west) (path picture bounding box.south west) -- (path picture bounding box.north east);
}}]

\node[draw,circle,cross,label=left:$g$] at (2,3) {};
\draw (1.2,3) arc (180:270:0.8);
\draw[decoration={markings, mark=at position 1 with {\arrow[scale=2]{>}}}, postaction={decorate}] (2,3) arc (0:180:0.4);
\draw[decoration={markings, mark=at position 0.6 with {\arrow[scale=2]{<}}}, postaction={decorate}] (2,3) -- (2,2.2);
\draw [fill] (2,2.2) circle [radius=0.07];

\node[draw,circle] at (2,1) {};
\draw[decoration={markings, mark=at position 0 with {\arrow[scale=2]{>}}}, postaction={decorate}] (2,1) circle (0.4) node[right=0.5cm] {$k$};
\draw [fill] (2,0.6) circle [radius=0.07];
\draw (2,0.6) -- (2,0) node[midway,right]{$1$};
\draw (2,0.6) arc (360:180:0.4) node[left] {$1$};
\draw plot [smooth,tension=1] coordinates{(1.2,0.6) (1.3,1.1) (1.6,1.4) (1.9,1.7) (2,2.2)};

\draw[decoration={markings, mark=at position 0 with {\arrow[scale=2]{>}}}, postaction={decorate}] (2,2) circle (2);
\draw [fill] (2,0) circle [radius=0.07];

\node[draw,circle,cross,label={[label distance=-0.1cm]180:$gk$}] at (7,1.4) {};
\draw (6.2,1.4) arc (180:270:0.8);
\draw[decoration={markings, mark=at position 0.9 with {\arrow[scale=2]{>}}}, postaction={decorate}] (7,1.4) arc (0:180:0.4);
\draw[decoration={markings, mark=at position 0.6 with {\arrow[scale=2]{<}}}, postaction={decorate}] (7,1.4) -- (7,0.6);
\draw [fill] (7,0.6) circle [radius=0.07];

\draw (7,0.6) -- (7,0) node[midway,right]{$1$};

\node[draw,circle] at (7,3) {};
\draw[decoration={markings, mark=at position 0 with {\arrow[scale=2]{>}}}, postaction={decorate}] (7,3) circle (0.4) node[right=0.5cm] {$k$};
\draw [fill] (7,2.6) circle [radius=0.07];

\draw (7,0.6) arc (360:270:0.3);
\draw[decoration={markings, mark=at position 0.85 with {\arrow[scale=2]{<}}}, postaction={decorate}] (6.7,0.3) arc (270:90:1) node[left=0.5]{$g$};
\draw (7,2.6) arc (360:270:0.3);

\draw[decoration={markings, mark=at position 0 with {\arrow[scale=2]{>}}}, postaction={decorate}] (7,2) circle (2) node [label={[label distance=-2.9cm]0:$\cong$}]{};
\draw [fill] (7,0) circle [radius=0.07];

\end{tikzpicture}
\caption{Axiom \eqref{eqax5}. To obtain the right figure from the left, the puncture with holonomy $k$ is pulled through the crosscap along the path with holonomy $g$.} \label{fig2}
\end{figure}

Finally, in the unoriented case we have cross-cap states $\theta_g\in \cA_{g^2}$, $g\notin G_0$. The state $\theta_g$, $g\notin G_0$, arises from a M\"obius strip with an oriented boundary and a base point on the boundary. The fundamental group of the M\"obius strip is isomorphic to $\ZZ$, where an orientation-reversing generator is fixed once the orientation of the boundary has been fixed. $\theta_g$ corresponds to a $G$-bundle whose holonomy along this generator is $g$.

The cross-cap states have the following properties:
\begin{equation}\label{eqax4}\alpha_{h\in G_0}(\theta_g)=\theta_{hgh^{-1}}\text{ and }\alpha_{h\notin G_0}(\theta_g)=\theta_{hg^{-1}h^{-1}}\end{equation}
\begin{equation}\label{eqax5}\theta_g\cdot\psi_k=\alpha_g(\psi_k)\cdot\theta_{gk}\text{ for all }\psi_k\in\cA_k.\end{equation}
\begin{equation}\label{eqax6}\sum_i\alpha_g(\xi^i_{gh})\xi_i^{gh}=\theta_g\cdot\theta_h.\end{equation}

\begin{figure}[ht]
\centering
\begin{tikzpicture}

\draw[decoration={markings, mark=at position 0.5 with {\arrow[scale=2]{>>}}}, postaction={decorate}] (0,0) -- (0,2);
\draw[decoration={markings, mark=at position 0.5 with {\arrow[scale=2]{>}}}, postaction={decorate}] (0,2) -- (2,2);
\draw[decoration={markings, mark=at position 0.5 with {\arrow[scale=2]{>>}}}, postaction={decorate}] (2,2) -- (2,0);
\draw[decoration={markings, mark=at position 0.5 with {\arrow[scale=2]{>}}}, postaction={decorate}] (2,0) -- (0,0);

\draw [fill] (1,1) circle [radius=0.07];

\draw[decoration={markings, mark=at position 0.75 with {\arrow[scale=2]{>}}}, postaction={decorate}] plot [smooth,tension=1] coordinates{(1.5,0) (1.3,0.7) (1,1) (0.7,1.3) (0.5,2)} node[below right=0.5cm]{$g$};

\draw[decoration={markings, mark=at position 0.5 with {\arrow[scale=2]{>>}}}, postaction={decorate}] (2.8,0) -- (2.8,2) node[midway,left=0.1cm]{$\#$};
\draw[decoration={markings, mark=at position 0.5 with {\arrow[scale=2]{>}}}, postaction={decorate}] (2.8,2) -- (4.8,2);
\draw[decoration={markings, mark=at position 0.5 with {\arrow[scale=2]{>>}}}, postaction={decorate}] (4.8,2) -- (4.8,0) node[midway,right=0.15cm]{$\cong$};
\draw[decoration={markings, mark=at position 0.5 with {\arrow[scale=2]{>}}}, postaction={decorate}] (4.8,0) -- (2.8,0);

\draw [fill] (3.8,1) circle [radius=0.07];

\draw[decoration={markings, mark=at position 0.75 with {\arrow[scale=2]{>}}}, postaction={decorate}] plot [smooth,tension=1] coordinates{(4.3,0) (4.1,0.7) (3.8,1) (3.5,1.3) (3.3,2)} node[below right=0.5cm]{$h$};

\draw[dotted,decoration={markings, mark=at position 0.5 with {\arrow[scale=2]{>>}}}, postaction={decorate}] (5.6,0) -- (5.6,2);
\draw[decoration={markings, mark=at position 0.5 with {\arrow[scale=2]{>}}}, postaction={decorate}] (5.6,2) -- (7.6,2);
\draw[decoration={markings, mark=at position 0.5 with {\arrow[scale=2]{>}}}, postaction={decorate}] (7.6,2) -- (9.6,2);
\draw[dotted,decoration={markings, mark=at position 0.5 with {\arrow[scale=2]{>>}}}, postaction={decorate}] (9.6,0) -- (9.6,2);
\draw[decoration={markings, mark=at position 0.5 with {\arrow[scale=2]{>}}}, postaction={decorate}] (9.6,0) -- (7.6,0);
\draw[decoration={markings, mark=at position 0.5 with {\arrow[scale=2]{>}}}, postaction={decorate}] (7.6,0) -- (5.6,0);
\draw (5.6,2) -- (7.6,0);
\draw (7.6,2) -- (9.6,0);
\draw [fill] (7.6,1) circle [radius=0.07];

\draw[decoration={markings, mark=at position 0.75 with {\arrow[scale=2]{>}}}, postaction={decorate}] plot [smooth,tension=1] coordinates{(8.1,0) (7.9,0.7) (7.6,1) (7.3,1.3) (7.1,2)} node[below right=0.5cm]{$g$};

\draw[decoration={markings, mark=at position 0.85 with {\arrow[scale=2]{>}}}, postaction={decorate}] plot [smooth,tension=1] coordinates{(5.6,0.5) (7.1,0.7) (7.6,1) (8.1,1.3) (9.6,1.5)} node[below left=0.15cm]{$gh$};

\end{tikzpicture}
\caption{Axiom \eqref{eqax6}. Two projective planes are punctured and sewed along their boundaries, the diagonal lines, to obtain their connected sum, the Klein bottle.} \label{fig3}
\end{figure}
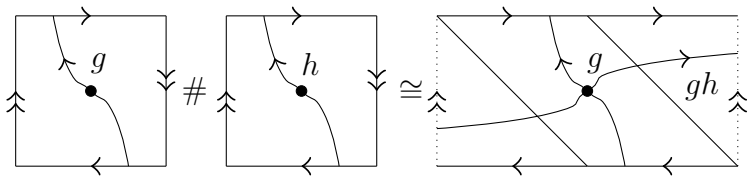

The first of these properties is illustrated in Figures \ref{fig1a} and \ref{fig1b}. The vectors $\theta_{h g^{(-1)} h^{-1}}$ and $\alpha_h(\theta_g)$ are defined by the two pictures which happen to be  related by an isotopy. The second property arises from an isotopy of the punctured M\"obius strip shown in Figure \ref{fig2}. The third property arises from the fact that a Klein bottle with two holes can be represented in two apparently different ways: as a cylinder with an orientation-reversing twist, or as a cylinder with an insertion of two cross-caps, see Figure \ref{fig3}.

We will call the data $(\cA, \eta, \alpha,\theta_g, g\notin G_0)$ algebraic TQFT data. In Appendix A we sketch a proof of 

\begin{prop}Unoriented equivariant $D=1$ TQFTs with symmetry $(G,\rho: G\rightarrow\ZZ_2)$ are in bijective correspondence with algebraic TQFT data $(\cA,\eta,\alpha,\theta_g, g\notin G_0)$. \label{propa}
\end{prop}
We have already explained how to assign algebraic TQFT data to any unoriented equivariant $D=1$ TQFT. The converse procedure is described in Appendix A. 

\subsection{Invertible unoriented equivariant $D=1$ TQFT}

Let us now specialize to the invertible case. For an invertible unoriented equivariant $D=1$ TQFT, the vector spaces $\cA_{g\in G_0}$ are one-dimensional. After fixing a basis $\{\ell_g\}_{g\in G_0}$ of $\cA$ so that $\eta(\ell_g,\ell_{g^{-1}})=1$, the algebraic TQFT data are determined by nonzero complex numbers $\theta(g)$, $g\notin G_0$, $z(h,k)$, $h,k\in G_0$, and $w(h,k)$, $h\notin G_0$, $k\in G_0$ defined as follows:
\begin{eqnarray*}
m_{k,l}(\ell_k,\ell_l)=b(k,l)\ell_{kl},&\quad&\theta_g=\theta(g)\ell_{g^2},\\\alpha_{h\in G_0}(\ell_k)=z(h,k)\ell_{hkh^{-1}},&\quad& \alpha_{h\notin G_0}(\ell_k)=w(h,k)\ell_{hk^{-1}h^{-1}}.\end{eqnarray*}
These numbers satisfy a number of identities following from the properties of algebraic TQFT data. 
\begin{prop}  Invertible unoriented equivariant $D=1$ TQFTs with symmetry $(G, \rho)$ are in bijective correspondence with elements of the $\rho$-twisted group cohomology $H^2(BG,\CC^*_\rho)\simeq H^2(BG,U(1)_\rho)$.\label{cocycle}\end{prop}

Twisted cohomology is the cohomology of the usual group cochain complex with respect to the $\rho$-twisted coboundary maps$$\delta^n_\rho:C^n(G,U(1))\rightarrow C^{n+1}(G,U(1)).$$ In degree $2$, the $\rho$-twisted cocycle condition reads
\begin{equation}
a(g,h)a(gh,k)=a(h,k)^{\rho(g)}a(g,hk)\label{cocycleeq}
\end{equation}

A proof of Proposition 2 is rather lengthy, see Appendix B. But the map in one direction,  from twisted group cohomology to the set of algebraic TQFT data, is easy to describe:\begin{eqnarray}b(k,l)&=&a(k,l)\label{f1}\\\theta(g)&=&a(g,g)\label{f2}\\z(h,k)&=&\frac{a(h,k)a(hk,h^{-1})}{a(h,h^{-1})}\label{f3}\\w(h,k)&=&\frac{a(h,k^{-1})a(hk^{-1},h^{-1})a(k,k^{-1})}{a(h,h^{-1})}\label{f4}\end{eqnarray}
To prove Proposition 2, we must show that these numbers satisfy the TQFT axioms \eqref{antiaut}-\eqref{eqax6} and that the map is injective and surjective. This is done in Appendix B.

\section*{Appendix A: Proof of Proposition \ref{propa}}

We have already shown that an unoriented equivariant $D=1$ TQFT has an underlying extended Turaev algebra $(\mathcal{A},\theta_g,\alpha_h)$. Oriented cobordisms and bundle isomorphisms constitute a $G_0$-crossed algebra $\mathcal{A}=\oplus_{g\in G_0}\mathcal{A}_g$, while crosscaps correspond to states $\theta_g\in\mathcal{A}_{g^2}$ and orientation-reversing homeomorphisms to algebra anti-automorphisms $\alpha_h:\mathcal{A}_g\rightarrow\mathcal{A}_{hg^{-1}h^{-1}}$. It remains to show the converse: that from each extended Turaev algebra we can construct an unoriented equivariant TQFT with this underlying algebra. We generalize the approaches of \cite{MooreSegal} and \cite{TuraevTurner} to unoriented equivariant theories.

We begin by defining the vector spaces assigned to simple objects $P_{[g],x,t}$ of the source category $\mathcal{C}$. To each circle $S$ equipped with principal $G$-bundle $P_{[g]}$, basepoint $x$, local trivialization $t:P_{[g]}|_x\rightarrow G$, and global trivialization of $o(S)\otimes\rho(P_{[g]})$, assign the vector space $\mathcal{H}(P_{[g],x,t})\cong\mathcal{A}_g$ where $g$ is the holonomy of $P_{[g]}$ around $S$ with respect to $x$ and $t$. Any object $E$ can be factored into simple objects $\sqcup_i P_{[g_i],x_i,t_i}$ and assigned a vector space $\mathcal{H}(E)\cong\otimes_i\mathcal{H}(P_{[g_i],x_i,t_i})$. It is clear that $\mathcal{H}(E)$ does not depend on the factorization of $E$.

Next we consider the linear maps assigned to morphisms of simple objects. One type of morphism $\tilde\alpha_k:P_{[g],x,t}\rightarrow P_{[g],y,s}$ arises from an isomorphism $f$ of the bundle $P_{[g]}$ where $(y,s)=(f(x),(f^{-1})^*t)$. Realized as its mapping cylinder, $f$ must have a global trivialization of $o(S\times I)\otimes\rho(f)$. Since $o(S\times I)$ is trivial, so must be $\rho(f)$, and so the holonomy of $P_{[g]}$ along a positive path from $(x,t)$ to $(y,s)$ is an element $k\in G_0$. We assign the linear map $\alpha_k:\mathcal{A}_g\rightarrow\mathcal{A}_{kgk^{-1}}$ to this morphism. The other type of morphism $\tilde\alpha_h:P_{[g],x,t}\rightarrow P_{[g^{-1}],y,s}$ arises from a bundle anti-isomorphism $P_{[g]}\rightarrow P_{[g^{-1}]}$ whose restriction to the base circle is not isotopic to the trivial homeomorphism. Since a bundle map of this type exchanges the sheets of $o(S)$, the holonomy of $P_{[g^{-1}]}$ from $(x,t)$ to $(y,s)$ is an element $h\notin G_0$. We assign the linear map $\alpha_h:\mathcal{A}_g\rightarrow\mathcal{A}_{hg^{-1}h^{-1}}$ to $\tilde\alpha_h$. This assignment is well defined for isomorphism classes of bundles, as the cylinder $\tilde\alpha_k\tilde\alpha_g$, related to $\tilde\alpha_k$ by a Dehn twist, is assigned the linear map $\alpha_k\alpha_g$, which equals $\alpha_k$ when restricted to $\mathcal{A}_g$ by \eqref{o4}.

Now we wish to define linear maps for cobordisms $(W,E_0,E_1)$. The strategy will be to decompose $W$ as a sequence of $n$ elementary cobordisms $(W^i,E_0^i,E_1^i)$, sewn along bundle (anti-)isomorphisms $s_i:E_1^i\rightarrow E_0^{i+1}$ with $E^0_0=E_0$ and $E^n_1=E_1$. After assigning a linear map to each $W^i$, we assign their composition $\tau(W)$ to $W$. We must then verify that $\tau(W)$ does not depend on the decomposition. Begin by considering the cobordism of base spaces $(N,M_0,M_1)$. By Sard's lemma, there exists a smooth function $f:N\rightarrow I$ such that $f^{-1}(0)=M_0$, $f^{-1}(1)=M_1$, and $f$ is Morse; that is, the gradient $df$ vanishes at finitely many critical points $x_i$, the Hessian $d^2f$ is a non-degenerate quadratic form at all $x_i$, and the critical values $c_i=f(x_i)$ are distinct and not equal to $0$ or $1$. The index $\text{ind}(x_i)$ is the number of negative eigenvalues of $d^2f$ at $x_i$. Choose $t_i\in I$ such that $0=t_0<c_1<t_1<\cdots<c_n<t_n=1$. By the implicit function theorem, each $M_{t_i}=f^{-1}(t_i)$ is a disjoint union of $m_i$ circles, and $\Sigma_i=f^{-1}([t_{i-1},t_i])$ is a cobordism from $M_{t_{i-1}}$ to $M_{t_i}$ with a single critical point. The classification of surfaces tells us that $\Sigma_i$ is homeomorphic to a disjoint union of cylinders and one of five possibilities: a cap, a pair-of-pants, their adjoints, and a twice-punctured real projective plane.

These spaces are base spaces for five classes of cobordisms $W$. Since any $G$-bundle over the disk is trivial, there is a unique cobordism over the cap, to which we assign the linear map $\eta:\mathcal{A}_1\rightarrow\mathbb{C}$. A $G$-bundle over the pair-of-pants, based and trivialized at the critical point, is almost determined by the holonomies $k$ and $l$ around the legs of the pants. We assign to it the linear map $m_{k,l}:\mathcal{A}_k\otimes\mathcal{A}_l\rightarrow\mathcal{A}_{kl}$. The orderings are related by conjugation $\alpha_l:\mathcal{A}_{kl}\rightarrow\mathcal{A}_{lk}$, and consistency requires that $m_{k,l}(\psi_k\otimes\psi_l)=\alpha_km_{l,k}(\psi_l\otimes\psi_k)$, which is enforced by the axioms \eqref{o4} and \eqref{o5} of the $G_0$-crossed algebra $\mathcal{A}$. The holonomies determine the bundle up to cylinders $\tilde\alpha_k$ sewn to the boundary circles, which were assigned maps $\alpha_k$ above. The next two maps are fixed by adjunction. The adjoint of $\eta$ distinguishes a state $\psi_\eta\in\mathcal{A}_1$ with the property that $\eta(\psi_\eta)=1$. The adjoint pair-of-pants is assigned a map $\Delta_{k,l}(\psi_{kl})=\sum_i\psi_{kl}\phi^i\otimes\phi_i$ where $\{\phi^i\}$ is a basis for $\mathcal{A}_l$ and $\{\phi_i\}$ is a dual basis for $\mathcal{A}_{l^{-1}}$. A $G$-bundle over the crosscap is specified (up to cylinders) by a holonomy $g\notin G_0$ around the orientation-reversing loop. We assign to it the linear map $\psi_k\mapsto m_{g^2,k}(\theta_g\otimes\psi_k)$, determined by the distinguished state $\theta_g\in\mathcal{A}_{g^2}$.

One may worry about a redundancy in the assignment of linear maps to composite cobordisms. Whenever an elementary cobordism $W^i$ and its sewing maps $s_{i-1}$ and $s_i$ can be modified in a way that preserves the composite cobordism $W$, consistency requires that $\tau(W)$ is also preserved. The map $s_i$ used to sew a cap or its adjoint into another cobordism does not affect the composite cobordism. The consistency of the algebraic description follows from the fact that $\alpha_k$ and $\alpha_h$ preserve $\eta$. Let $W^i$ be a pair-of-pants sewn along $s_{i-1}$ and $s_i$. Sewing instead along $(\tilde\alpha_k\otimes\tilde\alpha_k)\circ s_{i-1}$ and $s_i\circ\tilde\alpha_k^{-1}$ does not change $W$. Since $\alpha_k$ is an automorphism of $\mathcal{A}$, $\tau(W)$ is also preserved. Let $R$ be the bundle isomorphism that exchanges two circles. Then $(\tilde\alpha_h\otimes\tilde\alpha_h)\circ R\circ s_{i-1}$ and $s_i\circ\tilde\alpha_h^{-1}$ yield the same $W$. We require $\alpha_{h}^{-1}m(\alpha_h(\psi_l)\otimes\alpha_h(\psi_k))=m_{k,l}(\psi_k\otimes\psi_l)$, which is enforced by axiom \eqref{antiaut}. Let $(W^i,E^i_0,E^i_1)$ be a twice-punctured real projective plane with holonomy $g$ realized as a cobordism from $s_{i-1}:P_{[k]}\rightarrow E^i_0$ to $s_i:E^i_1\rightarrow P_{[g^2k]}$. There is a bundle isomorphism, covering a Dehn twist of the base space, between this cobordism and a twice-punctured real projective plane with holonomy $g^{-1}k^{-1}$ with sewing maps $s_{i-1}$ and $s_i\circ\tilde\alpha_g$. By axioms \eqref{antiaut} and \eqref{eqax5}, the consistency condition $\alpha_gm(\theta_{g^{-1}k^{-1}}\otimes\psi_k)=m_{g^2,k}(\theta_g\otimes\psi_k)$ is fulfilled. Now consider the M\"obius strip with holonomy $g\notin G_0$ constructed by sewing a cap into the twice-punctured real projective plane with holonomy $g$. Sewing this cobordism into another along $s_i$ yields the same composite cobordism related to the M\"obius strip with holonomy $hg^{-1}h^{-1}$ sewn along $\tilde\alpha_{h^{-1}}\circ s_i$ by a bundle isomorphism that covers a Y-homeomorphism of the base space. Axiom \eqref{eqax4} encodes this relation in the algebraic data.

The linear map $\tau(W)$ assigned to an arbitrary cobordism $W$ is given by the composition of maps assigned to its factors under Morse decomposition. It remains to show that $\tau(W)$ does not depend on the choice of Morse function. Any two Morse functions $f_0$ and $f_1$ are related by a smooth family of functions $f_s$ that are Morse at all but finitely many values of $s$. One possibility is that two critical points merge and annihilate for some $s$. Then $f_s$ has a degenerate critical point. This situation only occurs when deforming a pair-of-pants and an adjoint cap into a cylinder. For $\tau(W)$ to be consistent over the deformation, we require $m_{k,1}(\psi_k\otimes\psi_\eta)=\psi_k$. This condition is enforced by the axioms of $\mathcal{A}$. The remaining possibility is that two critical values coincide for some non-Morse value of $s$. We must check, for each composition W of two elementary cobordisms, that all factorizations give the same linear map. This situation occurs when both critical points have index $1$, in which case $W$ has Euler characteristic $\chi(W)=\sum_i (-1)^{\text{ind}(x_i)}=-2$. Hence $W$ is one of seven cobordisms: a genus zero oriented cobordism from three circles to one, its adjoint, a genus zero oriented cobordism from two circles to two, a twice-punctured torus from one circle to one, a crosscap-pants cobordism from two circles to one, its adjoint, and a twice-punctured Klein bottle from one circle to one.

The consistency of the first two cobordisms follows immediately from associativity of multiplication. The remaining two oriented conditions have been proven in Appendix A.3 of \cite{MooreSegal} and follow from the oriented axioms, notably \eqref{o6}. The next condition says that moving a crosscap from the ``torso'' to a leg of the pair-of-pants is a consistent deformation and also follows from associativity of multiplication. The Klein bottle has a decomposition as a pair-of-pants glued along its two legs to an adjoint pair-of-pants as well as a decomposition as a sphere with two crosscaps. The composite linear maps assigned to these realizations are equal to the others by axiom \eqref{eqax6}. We have assigned a linear map to each cobordism in terms of a Morse function $f$ and have seen that this map is independent of the choice of $f$. This completes the proof of Proposition \ref{propa}.

\section*{Appendix B: Proof of Proposition \ref{cocycle}}

Consider the map from $2$-cochains $a\in C^{n}(G,U(1))$ to TQFT data defined in \eqref{f1}-\eqref{f4}. If we restrict to the set $Z^2(G,U(1)_\rho)$ of $2$-cochains satisfying the $\rho$-twisted $2$-cocycle condition \eqref{cocycleeq}, we obtain a map $f$ from twisted cocycles to TQFT data. We will show that numbers in the image of $f$ satisfy the axioms \eqref{antiaut}-\eqref{eqax6}, and hence give rise to a consistent invertible UETQFT.

For an invertible theory, these axioms can be written as\begin{eqnarray*}w(h,kl)b(k,l)&=&w(h,k)w(h,l)b(hl^{-1}h^{-1},hk^{-1}h^{-1})\\w(h,g^2)\theta(g)&=&\theta(hg^{-1}h^{-1})\\ b(g^2,k)\theta(g)&=&b(gk^{-1}g^{-1},gkgk)w(g,k)\theta(gk)\\ b(g^2hg^{-1},gh)w(g,h^{-1}g^{-1})&=&\theta(g)\theta(h)b(g^2,h^2)b(h^{-1}g^{-1},gh)\end{eqnarray*}

It will be useful to impose a ``cyclic-symmetric gauge'' on the restriction of the cocycle $a$ to $G_0$:
$$
a(k,k^{-1})=1,\quad a(k,l)=a(l^{-1},k^{-1})^{-1},\quad \forall  k,l\in G_0.
$$
We also fix some $T\in G$ and impose the condition $a(k,T)=1$, $k\in G_0$.

Axiom \eqref{antiaut}:
\begin{eqnarray*}w(h,kl)b(k,l)&=&\frac{a(h,l^{-1}k^{-1})a(hl^{-1}k^{-1},h^{-1})a(kl,l^{-1}k^{-1})}{a(h,h^{-1})}a(k,l)\\&=&\frac{a(h,l^{-1}k^{-1})}{a(h,h^{-1})}\frac{a(hl^{-1},k^{-1}h^{-1})}{a(l^{-1},k^{-1})a(hl^{-1},k^{-1})a(k^{-1},h^{-1})}\\&=&\frac{a(hl^{-1},k^{-1}h^{-1})}{a(h,h^{-1})a(k^{-1},h^{-1})}a(h,l^{-1})\\&=&\frac{a(h,l^{-1})}{a(h,h^{-1})a(k^{-1},h^{-1})}\frac{a(hl^{-1}h^{-1},hk^{-1}h^{-1})a(h,k^{-1}h^{-1})}{a(hl^{-1}h^{-1},h)}\\&=&\frac{a(h,l^{-1})a(hl^{-1}h^{-1},hk^{-1}h^{-1})a(h,k^{-1}h^{-1})}{a(h,h^{-1})a(k^{-1},h^{-1})}a(h^{-1},h)a(hl^{-1},h^{-1})\\&=&\frac{a(h,l^{-1})a(hl^{-1},h^{-1})a(hl^{-1}h^{-1},hk^{-1}h^{-1})}{a(h,h^{-1})a(h,h^{-1})}a(h,k^{-1})a(hk^{-1},h^{-1})\\&=&w(h,k)w(h,l)b(hl^{-1}h^{-1},hk^{-1}h^{-1})\end{eqnarray*}

Axiom \eqref{eqax4}:
\begin{eqnarray*}w(h,g^2)\theta(g)&=&\frac{a(h,g^{-2})a(hg^{-2},h^{-1})a(g^2,g^{-2})}{a(h,h^{-1})}a(g,g)\\&=&\frac{a(hg^{-2},h^{-1})}{a(h,h^{-1})}a(hg^{-1},g^{-1})a(h,g^{-1})a(g,g)a(g^{-1},g^{-1})\\&=&\frac{a(h,g^{-1})a(g,g)a(g^{-1},g^{-1})}{a(h,h^{-1})}a(g^{-1},h^{-1})a(hg^{-1},g^{-1}h^{-1})\\&=&\frac{a(hg^{-1}h^{-1},hg^{-1}h^{-1})a(h,g^{-1})a(g,g)a(g^{-1},g^{-1})}{a(h,h^{-1})a(hgh^{-1},h)a(h,g^{-1}h^{-1})}a(g^{-1},h^{-1})\\&=&\frac{a(hg^{-1}h^{-1},hg^{-1}h^{-1})a(g,g)a(g^{-1},g^{-1})}{a(h,h^{-1})a(hgh^{-1},h)a(hg^{-1},h^{-1})}\frac{a(h,g^{-1})a(g^{-1},h^{-1})}{a(h,g^{-1})a(g^{-1},h^{-1})}\\&=&a(hg^{-1}h^{-1},hg^{-1}h^{-1})\\&=&\theta(hg^{-1}h^{-1})\end{eqnarray*}

Axiom \eqref{eqax5}:
\begin{eqnarray*}\theta(g)a(g^2,k)&=&\theta(g)a(g^2,k)\frac{a(g,k)a(g,k^{-1})a(k,k^{-1})}{a(g,g^{-1})}\frac{a(gk^{-1},k)}{a(g^{-1},gk)}\\&=&\theta(g)a(g^2,k)a(g,k)\frac{a(g,k^{-1})a(gk^{-1},g^{-1})}{a(g,g^{-1})}a(gk^{-1}g^{-1},gk)\\&=&\frac{a(g,k^{-1})a(gk^{-1},g^{-1})}{a(g,g^{-1})}a(g,gk)a(gk^{-1}g^{-1},gk)\\&=&a(gk^{-1}g^{-1},gkgk)\frac{a(g,k^{-1})a(gk^{-1},g^{-1})}{a(g,g^{-1})}a(gk,gk)\\&=&b(gk^{-1}g^{-1},gkgk)w(g,k)\theta(gk)\end{eqnarray*}

Axiom \eqref{eqax6}:
\begin{eqnarray*}b(g^2hg^{-1},gh)w(g,h^{-1}g^{-1})&=&a(g^2hg^{-1},gh)\frac{a(g,gh)a(g^2h,g^{-1})a(h^{-1}g^{-1},gh)}{a(g,g^{-1})}\\&=&\frac{a(g,gh)}{a(g,g^{-1})}\frac{a(g^2h,h)}{a(g^{-1},gh)}a(h^{-1}g^{-1},gh)\\&=&\frac{a(g^2,h)a(g,g)a(g,h)}{a(g,g^{-1})a(g^{-1},gh)}a(g^2h,h)a(h^{-1}g^{-1},gh)\\&=&\frac{a(g,g)a(g,h)}{a(g,g^{-1})a(g^{-1},gh)}a(g^2,h^2)a(h,h)a(h^{-1}g^{-1},gh)\\&=&a(g,g)a(h,h)a(g^2,h^2)a(h^{-1}g^{-1},gh)\\&=&\theta(g)\theta(h)b(g^2,h^2)b(h^{-1}g^{-1},gh)\end{eqnarray*}

We have shown that data in the image of $f$ define consistent invertible unoriented equivariant TQFTs. Both $Z^2(G,U(1)_\rho)$ and the set of invertible UETQFTs are groups, and it is easy to see that $f$ is a group homomorphism.

It remains to show that $f$ is injective and surjective. Let $(g,h,k)$ denote the twisted cocycle condition \eqref{cocycleeq}. We will construct a cocycle that solves \eqref{f1}-\eqref{f4}, an inverse to $f$.

Consider the twisted cocycle condition for $(k,T,T^{-1})$:
$$a(k,T)a(kT,T^{-1})=a(T,T^{-1}).$$Taking into account $a(k,T)=1$, we get
$a(kT,T^{-1})=a(T,T^{-1})$
This also implies
$a(Tk,T^{-1})=a(T,T^{-1}).$
So in this gauge we get
$w(T,k)=a(T,k^{-1})$. Next consider the twisted cocycle condition for  $(l,k,T)$:
$$a(l,k)a(lk,T)=a(l,kT)a(k,T).$$
Taking into account $a(k,T)=1$, we get
$a(l,kT)=a(l,k)$.
Since $T^{-2}\in G_0$, this implies
$a(k,T^{-1})=a(k,T^{-2}).$ Next consider the twisted cocycle condition for $(T,k,T^{-1})$:
$$a(T,k)a(Tk,T^{-1})a(k,T^{-1})=a(T,kT^{-1}).$$Using previous results, this is equivalent to$$a(T,kT^{-1})=a(k,T^{-2})a(T,T^{-1})a(T,k)$$Next consider the twisted cocycle condition for $(T,l,k)$:
$$a(Tl,k)a(T,l)a(l,k)=a(T,lk)$$Recall also that in our gauge $a(T,l)=w(T,l^{-1})$. Then
$$a(Tl,k)a(l,k)w(T,l^{-1})=w(T,k^{-1}l^{-1})$$
Since $\alpha_g\alpha_h=\alpha_{gh}$ and by axiom \eqref{antiaut}, we see
$$a(Tl,k)=w(T,k^{-1})a(TlT^{-1},TkT^{-1}).$$

We have determined the components of the twisted cocycle where one argument is in $G_0$ and the other is not. We have also determined $a(Tk,T^{-1})$ and $a(T,kT^{-1})$ up to a single term $a(T,T^{-1})$. We can determine $a(Tl,kT^{-1})$ by requiring that $a$ satisfies the twisted cocycle condition
$(T,l,kT^{-1})$:
$$a(Tl,kT^{-1})a(T,l)a(l,kT^{-1})=a(T,lkT^{-1})$$By construction, $a$ is a $2$-cochain that satisfies \eqref{f1}-\eqref{f4} as well as the $(k,T,T^{-1})$, $(l,k,T^{-1})$, $(T,k,T^{-1})$, $(T,l,k)$, $(l,k,m)$, and $(T,l,kT^{-1})$ cocycle conditions. The component $a(Tl,mT^{-1})$ is also determined by $(Tl,k,T^{-1})$, and equality of the two expressions must hold if $a$ is a cocycle:
$$a(Tl,kT^{-1})=a(Tlk,T^{-1})a(Tl,k)a(k,T^{-1})$$
In the above expression, apply the $(T,lk,T^{-1})$ condition to the first term to obtain $\frac{a(T,lkT^{-1})}{a(T,lk)a(lk,T^{-1})}$. Hit the second term with $(T,l,k)$ to obtain $\frac{a(T,lk)}{a(l,k)a(T,l)}$. Hit $a(lk,T^{-1})$ with $(l,k,T^{-1})$ to get $\frac{a(l,kT^{-1})a(k,T^{-1})}{a(l,k)}$. After cancellation, we are left with the first expression for $a(Tl,kT^{-1})$. 

To see injectivity of $f$, consider the trivial TQFT with $b$, $w$, $\theta$ trivial. The cocycle solution has $a(k,l)=1$ and $a(k,lT)=1$. We have $a(Tl,k)=\frac{w(T,k^{-1}l^{-1})}{a(l,k)w(T,l^{-1})}=1$ as well as $$a(Tl,kT^{-1})=a(Tlk,T^{-1})a(Tl,k)a(k,T^{-1})=\theta(T^{-1})=1$$so the only the trivial cocycle corresponds to the trivial theory.

It remains to show that $a$ satisfies the cocycle condition for all possible combinations of arguments; in particular, we must show the $(k,l,mT)$, $(kT,l,m)$, $(k,lT,m)$, $(k,lT,mT)$, $(kT,l,mT)$, $(kT,lT,m)$, and $(kT,lT,mT)$ conditions. Consider the first condition:$$a(k,l)a(kl,mT)=a(l,mT)a(l,kmT)$$Since $a(k,lT)=a(k,l)$ for all $k,l\in G_0$ in our gauge, this follows from the $G_0$ cocycle condition. Now consider the third:$$a(kT,l)a(kTl,m)a(l,m)=a(lT,km)$$Apply the $(T,k,l)$ condition to the first term to get$\frac{a(T,kl)}{a(k,l)a(T,k)}$, the $(T,kl,m)$ condition to the second term to get $\frac{a(T,klm)}{a(kl,m)a(T,kl)}$, and the $(T,k,lm)$ condition to the third term to get $\frac{a(T,lkm)a(T,m)}{a(k,lm)}$. The desired condition is reduced to a known condition.

Now consider $(Tk,l,mT^{-1})$:$$a(Tk,l)a(Tkl,mT^{-1})a(l,mT^{-1})=a(Tk,lmT^{-1})$$ The first term becomes $\frac{a(T,kl)}{a(T,k)a(k,l)}$ after $(T,k,l)$, the second $\frac{a(T,klmT^{-1})}{a(T,kl)a(kl,mT^{-1})}$ after $(T,kl,mT^{-1})$, the third $(a(k,l)a(kl,mT^{-1}))^{-1}$ after $(l,m,T^{-1})$, and the fourth $\frac{a(T,klmT^{-1}}{a(T,k)a(k,lmT^{-1})}$ after $(T,k,lmT^{-1})$. Everything cancels.

Since $a(kT,T^{-1})=a(T,T^{-1})$, we get the $(l,lT,T^{-1})$ condition by applying $(kl,T,T^{-1})$ to $a(klT,T^{-1})$. Then $(k,lT,mT^{-1})$ reads$$a(k,lT)a(klt,mT^{-1})=a(lT,mT^{-1})a(k,lTmT^{-1})$$The last term is just $a(k,lTm)$ in our gauge and becomes $\frac{a(k,lT)a(klT,m)}{a(lT,m)}$ after $(k,lT,m)$. $a(klT,mT^{-1})$ becomes $\frac{a(klTm,T^{-1})a(klT,m)}{a(m,T^{-1})}$ after $(klT,m,T^{-1})$, and $a(lT,mT^{-1})$ becomes $\frac{a(lTm,T^{-1})a(lT,m)}{a(m,T^{-1})}$ after $(lT,m,T^{-1})$. We have seen that $a(klTm,T^{-1})=a(T,T^{-1})=a(lTm,T^{-1})$ so we are done.

The condition $(k,lT,T^{-1})$ is shown by noting that $a(k,lT)=a(k,l)$ and $a(klT,T^{-1})=a(T,T^{-1})=a(lT,T^{-1})$. Consider the $(k,Tl,mT^{-1})$ condition:$$a(k,Tl)a(kTl,mT^{-1})=a(Tl,mT^{-1})a(k,TlmT^{-1})$$Hit the second term with $(kTl,m,T^{-1})$ to get $a(kTlm,T^{-1})a(m,T^{-1})a(kTl,m)$ and the fourth term with $(k,Tlm,T^{-1})$ to get $\frac{a(kTlm,T^{-1})a(k,Tlm)}{a(Tlm,T^{-1})}$. Then $a(kTl,m)$ becomes $\frac{a(k,Tlm)a(Tl,m)}{a(k,TL)}$ by $(k,Tl,m)$ and $a(Tlm,T^{-1})$ becomes $\frac{a(Tl,m)a(m,T^{-1})}{a(Tl,mT^{-1})}$ by $(T,lm,T^{-1})$.

Consider $(T,T,T)$:$$a(T^2,T)a(T,T)a(T,T)=a(T,T^2)$$The first term vanishes, and we are left with $\theta(T)^2=w(T,T^{-2})$, which is true by axiom \eqref{eqax6} with $g=h=T$.

Consider $(lT^{-1},T,T)$: $$a(lT^{-1},T)a(T,T)=a(lT^{-1},T^2)$$ The first term is just $\frac{a(T^{-1},T)}{a(l,T^{-1})}$ by $(l,T^{-1},T)$. The third is $\frac{a(T^{-1},T^2)}{a(l,T^{-1})}$. The condition then follows from $(T,T,T)$. Consider $(T,mT^{-1},T)$:$$a(T,mT^{-1})=a(T,m)a(mT^{-1},T)$$The first term becomes $a(Tm,T^{-1})a(T,m)a(m,T^{-1})$ by $(T,m,T^{-1})$ and the second becomes $\frac{a(T^{-1},T)}{a(m,T^{-1})}$. We are left with $a(Tm,T^{-1})a(T^{-1},T)=1$. This is $\theta(T^{-1})\theta(T)$, which vanishes by axiom \eqref{eqax6}. Now consider $(kT,lT^{-1},T)$:$$a(kT,lT^{-1})a(lT^{-1})=a(kT,l)$$The first term is $a(kTl,T^{-1})a(kT,l)a(l,T^{-1})$ by $(kT,l,T^{-1})$ and the second is $\frac{a(T^{-1},T)}{a(l,T^{-1})}$ by $(l,T^{-1},T)$. We are left with $a(kTl,T^{-1})a(T^{-1},T)=1$ which holds as before.

Start with the $(T^{-1},T,mT^{-1})$ cocycle condition:
$$a(T^{-1},T)a(T,mT^{-1})=a(T^{-1},TmT^{-1})$$Apply $(T,m,T^{-1})$ to the third term. It becomes $a(Tm,T^{-1})a(T,m)a(m,T^{-1})$. Note that
$a(T,m)=\frac{w(T,m^{-1})a(T,T^{-1})}{a(Tm,T^{-1})}$
and that
$a(T^{-1},TmT^{-1})=\frac{w(T^{-1},Tm^{-1}T^{-1})a(T^{-1},T)}{a(mT^{-1},T)}$.
By $(m,T,T^{-1})$, we have
$a(mT^{-1},T) = \frac{a(T^{-1},T)}{a(m,T^{-1})}$
The first equation becomes
$$w(T,m^{-1})=w(T^{-1},Tm^{-1}T^{-1})$$
Since $\alpha_T\alpha_{T^{-1}}=1$, this becomes
$w(T,m^{-1})w(T,m)=1$
which is true by axiom \eqref{antiaut}. This proves the $(T^{-1},T,mT^{-1})$ cocycle condition.

Now consider the $(lT^{-1},T,m)$ condition: $$a(lT^{-1},T)a(l,m)a(T,m)=a(lT^{-1},Tm)$$Hit the first term with $(l,T^{-1},T)$ to get $\frac{a(T^{-1},T)}{a(l,T^{-1})}$ and the fourth term with $(l,T^{-1},Tm)$ to get $\frac{a(l,m)a(T^{-1},Tm)}{a(l,T^{-1})}$. Apply the new result $(T^{-1},T,m)$ to $a(T^{-1},Tm)$ to get $a(T,m)a(T^{-1},T)$. Everything cancels. This proves $(lT^{-1},T,m)$.

Now consider the $(lT,kT,m)$ condition: $$a(lT,kT)a(lTkT,m)a(kT,m)=a(lT,kTm)$$Hit the first term with $(lT,k,T)$, the second term with the new result $(lTk,T,m)$, the third term with $(k,T,m)$, and the fourth term with $(lT,k,Tm)$. Everything cancels.

Finally, check $(kT,T^{-1}l,mT)$:$$a(kT,T^{-1}l)a(kl,m)a(T^{-1}l,mT)=a(kT,T^{-1}lmT)$$The last term becomes $a(kT,T^{-1}lm)a(T^{-1}lm,T)$ by $(kT,T^{-1}lm,T)$. $a(kT,T^{-1}lm)$ becomes $a(kl,m)a(kT,T^{-1}l)a(T^{-1}l,m)$ by $(kT,T^{-1}l,m)$ and $a(T^{-1}lm,T)$ becomes $\frac{a(T^{-1}l,mT)}{a(T^{-1}l,m)}$ by $(T^{-1}l,m,T)$. Everything cancels, proving the last cocycle condition $(kT,lT,mT)$.

This proves that each invertible unoriented equivariant TQFT arises from a twisted 2-cocycle. Since this twisted 2-cocycle gives an inverse to $f$, we have shown that $f$ is surjective. This completes the proof of Proposition \ref{cocycle}.

\end{document}